\documentclass[10pt,technote]{IEEEtran}

\usepackage{amsmath,amssymb,bm,amsthm,mathtools}
\usepackage{graphicx}
\usepackage{cite}
\usepackage{algorithm}
\usepackage{algorithmic}
\usepackage{url}
\usepackage{xcolor}

\newtheorem{lemma}{Lemma}

\newcommand{\Tr}{\operatorname{Tr}}
\newcommand{\diag}{\operatorname{diag}}
\newcommand{\blkdiag}{\operatorname{blkdiag}}
\newcommand{\vect}{\operatorname{vec}}
\newcommand{\Real}{\operatorname{Re}}
\newcommand{\Imag}{\operatorname{Im}}
\newcommand{\CN}{\mathcal{CN}}
\newcommand{\cC}{\mathbb{C}}
\newcommand{\cR}{\mathbb{R}}
\newcommand{\cK}{\mathcal{K}}
\newcommand{\cJ}{\mathcal{J}}

\newcommand{\cX}{\mathcal{X}}
\newcommand{\cQ}{\mathcal{Q}}

\newcommand{\jmathc}{\mathrm{j}}
\newcommand{\norm}[1]{\left\lVert #1\right\rVert}
\newcommand{\absval}[1]{\left\lvert #1\right\rvert}

\newcommand{\safeincludegraphics}[2][]{%
  \IfFileExists{#2}{\includegraphics[#1]{#2}}{%
  \fbox{\parbox[c][1.42in][c]{0.92\linewidth}{\centering
  \footnotesize Figure file \texttt{\detokenize{#2}} is not available.}}}}

\title{Pinching Antenna Systems (PASS) for Air-Ground Integrated Sensing and Communications (ISAC)}
\author{
{Chen~Chaoran,~\IEEEmembership{Graduate~Student~Member,~IEEE}},
{Li~Haochen,~\IEEEmembership{Member,~IEEE}},
{Zhu~Jincheng,~\IEEEmembership{Graduate~Student~Member,~IEEE}},
{Pan~Zhiwen,~\IEEEmembership{Member,~IEEE}},
{Mark F. Flanagan,~\IEEEmembership{Senior Member,~IEEE}}

\thanks{
Chen~Chaoran, Zhu~Jincheng, and Pan~Zhiwen are with the National Mobile Communications Research Laboratory, Southeast University, Nanjing 210096, China (e-mail: chenchaoran@seu.edu.cn; jincheng.zhu@seu.edu.cn; pzw@seu.edu.cn). Li~Haochen is with the College of Electronic and Information Engineering, Nanjing University of Aeronautics and Astronautics, Nanjing 211106, China (haochen.li@nuaa.edu.cn).  M.~F.~Flanagan is with the School of Electrical and Electronic Engineering, University College Dublin, Belfield, Dublin 4, Ireland (e-mail: mark.flanagan@ieee.org).}
}

\begin{document}
\maketitle

\begin{abstract}
This paper proposes a multi-user single-target air-ground integrated sensing and communication (ISAC) system, in which a ground pinching-antenna system (PASS) transmits communication and sensing signals and an unmanned aerial vehicle (UAV) receives the target echo.
The communication and sensing beamforming vectors, pinching antenna positions, and UAV position are jointly optimized to minimize the weighted Cram\'er-Rao bound (CRB).
A penalty dual decomposition (PDD) algorithm is developed for the resulting nonconvex problem. 
The auxiliary variable and beamforming subproblems admit globally optimal solutions in semi-closed form, while the positions of the pinching antennas and the UAV are updated by successive convex approximation (SCA). 
Under mild conditions, every accumulation point of the sequence generated by our proposed algorithm is a Karush-Kuhn-Tucker (KKT) point of the reformulated problem.
Numerical results show that the proposed PASS-UAV design achieves a lower weighted CRB than conventional multiple-input multiple-output (MIMO) and PASS benchmarks. 
\end{abstract}

\vspace{-0.1cm}

\section{Introduction}

Integrated sensing and communication (ISAC) has attracted significant attention as a promising technology for future wireless networks~\cite{zhang2024input}. 
However, the fixed antenna positions of conventional antenna arrays limit their flexibility in adapting the communication and sensing channels. 
Pinching-antenna systems (PASS) offer a new degree of freedom by allowing pinching antennas (PAs) to move along dielectric waveguides, thereby enabling flexible channel reconfiguration for ISAC design~\cite{11434944}.

Existing PASS sensing designs mainly employ fixed terrestrial sensing receivers~\cite{jiang2026pinching,li2026pinching,song2026pinching,guo2025learning,hu2025pinching}. 
Moreover, in existing PASS-enabled low-altitude networks, unmanned aerial vehicles (UAVs) are generally treated as communication terminals or sensing targets~\cite{guo2025learning,hu2025pinching}, overlooking the potential of exploiting their mobility to actively configure the sensing receive geometry. 
A UAV receiver can adjust both its distance from the target and the angle of arrival of the echo, while the movable PAs control the transmit channel.
This motivates a joint PASS transmitter and UAV receiver design.

Motivated by this observation, we propose a PASS-assisted air-ground ISAC system in which a ground PASS serves multiple communication users and senses one ground target, while a UAV collects the target echo. 
The communication and sensing beamforming vectors, PA positions, and UAV position are jointly optimized under communication rate, transmit power, PA deployment, and UAV deployment constraints. 
The problem is challenging because the PASS channels depend nonlinearly on the PA positions, the communication rate constraints are nonconvex, and the Cram\'{e}r-Rao bound (CRB) couples the transmit design with the UAV receive geometry.

The main contributions of this work are summarized as follows.
First, we establish the signal and sensing models for the proposed PASS-UAV ISAC system and formulate the joint beamforming, PA position, and UAV position design problem. 
Second, we develop a penalty dual decomposition (PDD) algorithm~\cite{shi2020pdd} in which the auxiliary variable and beamforming subproblems admit globally optimal solutions, while the PA and UAV positions are updated by successive convex approximation (SCA) using quadratic upper bounds and backtracking. 
Third, numerical results demonstrate the gains of the proposed joint design over conventional multiple-input multiple-output (MIMO) and PASS benchmarks.\vspace{-0.1cm}

\section{System Model and Problem Formulation}\label{sec:system_model}
\begin{figure}[t]
\centering
\safeincludegraphics[width=0.7\linewidth]{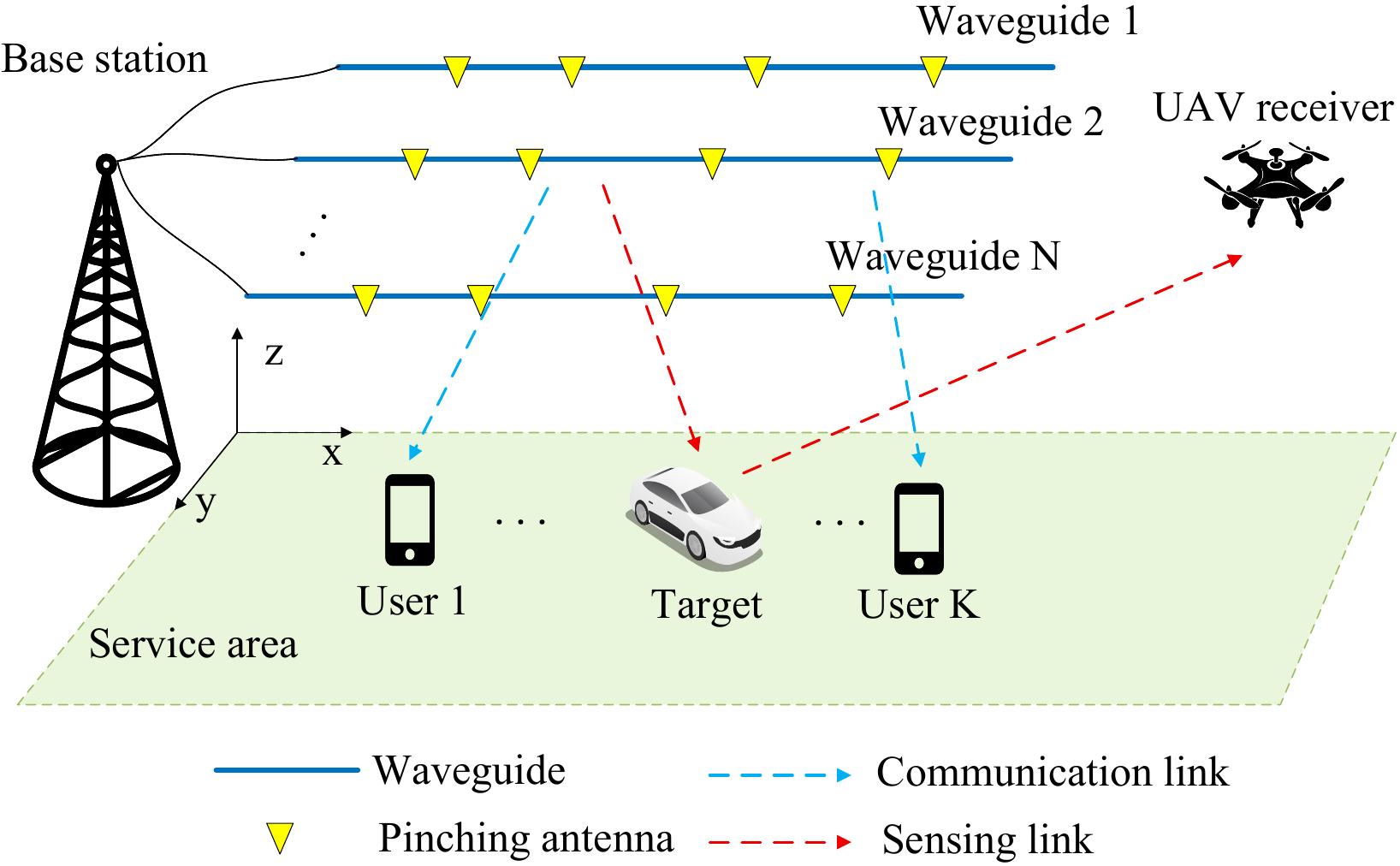}
\caption{PASS-assisted air-ground ISAC with a UAV sensing receiver.}\vspace{-0.2cm}
\label{fig:scenario}
\end{figure}

As shown in Fig.~\ref{fig:scenario}, a ground base station (BS) equipped with a PASS serves $K$ single-antenna users and senses one ground target. 
The PASS contains $N$ parallel waveguides, each equipped with $M$ movable PAs, and a UAV is deployed to collect the sensing echo.
Let $\cK=\{1,\ldots,K\}$ denote the set of communication users and $D\triangleq K+1$ denote the total number of transmit streams.

User $k$ is located at $\bm l_{\mathrm{c},k}=[x_{\mathrm{c},k},y_{\mathrm{c},k},0]^{\mathrm{T}}$. 
Next, we consider sensing a target whose location within an area of interest is unknown, while its discrete prior distribution is assumed to be available to the system based on a previous estimation (i.e., tracking phase).
Specifically, the target is located at $\bm l_{\mathrm{s},j}=[x_{\mathrm{s},j},y_{\mathrm{s},j},0]^{\mathrm{T}}$, $j\in\cJ=\{1,\ldots,J\}$, with probability $\omega_j>0$, where $\sum_{j\in\cJ}\omega_j=1$.
The waveguides are oriented along the $x$-axis at height $d_{\mathrm{h}}$, with the feed point of waveguide $n$ located at $\bm l_n=[x_{\mathrm{f}},y_n,d_{\mathrm{h}}]^{\mathrm{T}}$. 
The $m$-th PA on the $n$-th waveguide is located at $\bm l_{m,n}=[x_{m,n},y_n,d_{\mathrm{h}}]^{\mathrm{T}}$ and satisfies $x_{\mathrm{f}}\leq x_{1,n}$, $x_{M,n}\leq x_{\mathrm{f}}+L$, and $x_{m,n}-x_{m-1,n}\geq\delta$ for $m=2,\ldots,M$. 
Here, $L$ and $\delta$ denote the waveguide length and minimum PA spacing. Let $\bm X=[x_{m,n}]\in\cR^{M\times N}$ and denote its feasible set by $\cX$.\vspace{-0.1cm}

\subsection{PASS Channel Model}
The PASS channel consists of the wireless propagation channels and the in-waveguide channels. 
Let $\bm h_{k,n}^{\mathrm{c}}\in\cC^M$ and $\bm h_{j,n}^{\mathrm{s}}\in\cC^M$ denote the wireless channels from the PAs on waveguide $n$ to user $k$ and to potential target position $j$, respectively. Their $m$-th elements are $[\bm h_{k,n}^{\mathrm{c}}]_m=e^{-\jmathc\kappa_{\mathrm{c}} r_{k,m,n}^{\mathrm{c}}}/(2\kappa_{\mathrm{c}} r_{k,m,n}^{\mathrm{c}})$ and $[\bm h_{j,n}^{\mathrm{s}}]_m=e^{-\jmathc\kappa_{\mathrm{c}} r_{j,m,n}^{\mathrm{s}}}/(2\kappa_{\mathrm{c}} r_{j,m,n}^{\mathrm{s}})$, where $r_{k,m,n}^{\mathrm{c}}=\norm{\bm l_{m,n}-\bm l_{\mathrm{c},k}}_2$ and $r_{j,m,n}^{\mathrm{s}}=\norm{\bm l_{m,n}-\bm l_{\mathrm{s},j}}_2$.
Here, $\kappa_{\mathrm{c}}=2\pi/\lambda_{\mathrm{c}}$, where $\lambda_{\mathrm{c}}=c/f_{\mathrm{c}}$ is the wavelength corresponding to carrier frequency $f_{\mathrm{c}}$ and $c$ is the speed of light. 
Then, the overall wireless channels can be expressed as
\begin{equation}
\begin{aligned}
\bm h_k^\mathrm{c}(\bm X)&=\left[(\bm h_{k,1}^\mathrm{c})^\mathrm{T},\ldots,(\bm h_{k,N}^\mathrm{c})^\mathrm{T}\right]^\mathrm{T}\in\cC^{MN},\\
\bm h_j^\mathrm{s}(\bm X)&=\left[(\bm h_{j,1}^\mathrm{s})^\mathrm{T},\ldots,(\bm h_{j,N}^\mathrm{s})^\mathrm{T}\right]^\mathrm{T}\in\cC^{MN}.
\end{aligned}
\label{eq:wireless_channel_stacking}
\end{equation}
The overall in-waveguide channel between all feed points of the PASS and all PAs is in the form of a block-diagonal matrix, which can be expressed as
\begin{equation}
\bm F(\bm X)=\blkdiag(\bm f_1^{\mathrm{T}},\ldots,\bm f_N^{\mathrm{T}})\in\cC^{N\times MN},
\label{eq:guided_channel}
\end{equation}
where $\bm f_n=M^{-1/2}[e^{-\jmathc\kappa_{\mathrm{g}}(x_{1,n}-x_{\mathrm{f}})},\ldots,e^{-\jmathc\kappa_{\mathrm{g}}(x_{M,n}-x_{\mathrm{f}})}]^{\mathrm{T}}$, $\kappa_{\mathrm{g}}=2\pi/\lambda_{\mathrm{g}}$, $\lambda_{\mathrm{g}}=\lambda_{\mathrm{c}}/n_{\mathrm{e}}$, and $n_{\mathrm{e}}$ is the effective refractive index of the dielectric waveguides. 
The resulting overall PASS channels can be expressed as
\begin{equation}
\begin{aligned}
\bm h_{\mathrm{c},k}(\bm X)&=\bm F(\bm X)\bm h_k^{\mathrm{c}}(\bm X)\in\cC^N,\quad k\in\cK,\\
\bm h_{\mathrm{s},j}(\bm X)&=\bm F(\bm X)\bm h_j^{\mathrm{s}}(\bm X)\in\cC^N,\quad j\in\cJ.
\end{aligned}
\label{eq:overall_PASS_channels}\vspace{-0.1cm}
\end{equation}

\subsection{Signal Model}
At symbol interval $\ell$, the BS transmits an ISAC signal consisting of $K$ communication streams and one dedicated sensing stream, given by
\begin{equation}\label{eq:transmit_signal}
\bm u[\ell]=\sum\nolimits_{k=1}^{K}\bm w_k s_k[\ell]+\bm v s_0[\ell],
\end{equation}
where $\bm w_k\in\cC^N$ is the communication beamforming vector for user $k$, $s_k[\ell]$ is the information-bearing symbol intended for user $k$, $\bm v\in\cC^N$ is the sensing beamforming vector, and $s_0[\ell]$ is the dedicated sensing symbol. 
Assume all signals have unit energy and are mutually independent.
Defining the beamforming matrix $\bm B=[\bm w_1,\ldots,\bm w_K,\bm v]\in\cC^{N\times D}$, the transmit covariance is $\bm R_u=\mathbb E\{\bm u[\ell]\bm u^{\mathrm{H}}[\ell]\}=\bm B\bm B^{\mathrm{H}}$. 
Over $T_{\mathrm{s}}$ sensing symbols, let $\bm S=[\bm u[1],\ldots,\bm u[T_{\mathrm{s}}]]$. 
For sufficiently large $T_{\mathrm{s}}$, we use the approximation $T_{\mathrm{s}}^{-1}\bm S\bm S^{\mathrm{H}}\approx\bm R_u=\bm B\bm B^{\mathrm{H}}$.

\subsection{Communication Model}
The received signal at user $k$ is $y_k[\ell]=\bm h_{\mathrm{c},k}^{\mathrm{H}}(\bm X)\bm u[\ell]+n_k[\ell]$, where $n_k[\ell]\sim\CN(0,\sigma_{\mathrm{c}}^2)$ is the additive white Gaussian noise with $\sigma_{\mathrm{c}}^2$ being the noise power.
The resulting signal-to-interference-plus-noise ratio (SINR) of user $k$ is
\begin{equation}
\gamma_k=\frac{|\bm h_{\mathrm{c},k}^{\mathrm{H}}(\bm X)\bm w_k|^2}{\sum_{i\neq k}|\bm h_{\mathrm{c},k}^{\mathrm{H}}(\bm X)\bm w_i|^2+|\bm h_{\mathrm{c},k}^{\mathrm{H}}(\bm X)\bm v|^2+\sigma_{\mathrm{c}}^2},
\label{eq:sinr}
\end{equation}
and the achievable rate of user $k$ is $R_k=\log_2(1+\gamma_k)$.\vspace{-0.1cm}

\subsection{Sensing Model}
The UAV is equipped with a $U$-element uniform linear array (ULA) oriented along the $x$-axis, whose center is located at $\bm q_{\rm U}=[x_{\rm U},y_{\rm U},h_{\rm U}]^{\mathrm{T}}$, where $h_{\rm U}$ is fixed and $\bm q_{\rm U}\in\cQ\triangleq\{[x,y,h_{\rm U}]^{\mathrm{T}}:(x,y)\in\cQ_{\mathrm{h}}\}$ with $\cQ_{\mathrm{h}}=[x_{\min},x_{\max}]\times[y_{\min},y_{\max}]$. Let $d_{j,{\rm U}}=\norm{\bm l_{\mathrm{s},j}-\bm q_{\rm U}}_2$ denote the target-to-UAV distance and let $\theta_j$ denote the corresponding angle of arrival (AoA), with $\sin\theta_j=(x_{\mathrm{s},j}-x_{\rm U})/d_{j,{\rm U}}$. 
Under the far-field model, the steering vector is $\bm a_{\rm U}(\theta_j)=[1,e^{-\jmathc\kappa_{\rm U}\sin\theta_j},\ldots,e^{-\jmathc\kappa_{\rm U}(U-1)\sin\theta_j}]^{\mathrm{T}}$, where $\kappa_{\rm U}=2\pi d_{\rm U}/\lambda_{\mathrm{c}}$ and $d_{\rm U}=\lambda_{\mathrm{c}}/2$ is the antenna spacing of the UAV ULA.

Assuming time and frequency synchronization during each sensing block, as well as perfect suppression of the direct PASS-to-UAV signal, the received signal corresponding to potential target location $\bm l_{\mathrm{s},j}$ at snapshot $\ell$ is given by
\begin{equation}
\bm y_{{\rm U},j}[\ell]=\eta_j\bm a_{\rm U}(\theta_j)\bm h_{\mathrm{s},j}^{\mathrm{H}}(\bm X)\bm u[\ell]+\bm n_{{\rm U},j}[\ell],\quad j\in\cJ,
\label{eq:uav_snapshot}
\end{equation}
where $\eta_j=\alpha\beta_{j,{\rm U}}$, $\beta_{j,{\rm U}}=\lambda_{\mathrm{c}}/(4\pi d_{j,{\rm U}})$ is the target-to-UAV path-loss coefficient under potential position $j$, $\alpha$ denotes the complex bistatic target scattering coefficient, and $\bm n_{{\rm U},j}[\ell]\sim\CN(\bm0,\sigma_{\mathrm{s}}^2\bm I_U)$. 
It is shown in Appendix~\ref{app:AoA_CRB} that the CRB for estimating $\theta_j$ is given by
\begin{equation}
C_{\theta,j}=\frac{\chi\zeta_j(\bm q_{\rm U})}{\bm h_{\mathrm{s},j}^{\mathrm{H}}(\bm X)\bm B\bm B^{\mathrm{H}}\bm h_{\mathrm{s},j}(\bm X)},\quad j\in\cJ,
\label{eq:conditional_AoA_CRB}
\end{equation}
where $\zeta_j(\bm q_{\rm U})=d_{j,{\rm U}}^2/\cos^2\theta_j$, and $\chi=96\pi^2\sigma_{\mathrm{s}}^2/[T_{\mathrm{s}}|\alpha|^2\lambda_{\mathrm{c}}^2\kappa_{\rm U}^2U(U^2-1)]$ is a positive constant independent of $j$ and the optimization variables.
The overall sensing performance is characterized by the weighted CRB
\begin{equation}
\overline C_{\theta}(\bm B,\bm X,\bm q_{\rm U})
\triangleq\sum\nolimits_{j\in\cJ}\omega_j C_{\theta,j}
\label{eq:weighted_AoA_CRB}\vspace{-0.1cm}
\end{equation}

\subsection{Problem Formulation}
The beamforming vectors, PA positions, and UAV position are jointly optimized to minimize the weighted CRB by solving
\begin{subequations}\label{prob:P1}
\begin{align}
\min_{\bm B,\bm X,\bm q_{\rm U}}\quad&\overline C_\theta\label{prob:P1_obj}\\
\mathrm{s.t.}\quad&R_k\geq R_k^{\min},\quad k\in\cK,\label{prob:P1_qos}\\
&\norm{\bm B}_{\mathrm{F}}^2\leq P_{\max},\label{prob:P1_power}\\
&\bm X\in\cX,\label{prob:P1_X}\\
&\bm q_{\rm U}\in\cQ.\label{prob:P1_Q}
\end{align}
\end{subequations}
Here, $R_k^{\min}$ denotes the communication rate requirement of user $k$, and $P_{\max}$ denotes the BS transmit power budget. 
Problem~\eqref{prob:P1} is nonconvex due to the nonconvex CRB terms in the objective and the nonconvex fractional SINR constraints. Moreover, the beamforming matrix, PA positions, and UAV position are tightly coupled in the CRB objective, with the beamforming matrix and PA positions further coupled in the communication constraints, making the problem difficult to solve directly.\vspace{-0.1cm}

\section{Proposed PDD Algorithm}\label{sec:proposed_algorithm}
\subsection{Transformation of Problem \eqref{prob:P1}}

To tackle problem~\eqref{prob:P1}, introduce the auxiliary variables
\begin{equation}
\bm A_{\mathrm{c}}=\bm H_{\mathrm{c}}^{\mathrm{H}}(\bm X)\bm B\in\cC^{K\!\times\!D},
\bm A_{\mathrm{s}}=\bm H_{\mathrm{s}}^{\mathrm{H}}(\bm X)\bm B\in\cC^{J\!\times\!D},
\label{eq:auxiliary_equalities}
\end{equation}
where $\bm H_{\mathrm{c}}(\bm X)=[\bm h_{\mathrm{c},1}(\bm X),\ldots,\bm h_{\mathrm{c},K}(\bm X)]$ and $\bm H_{\mathrm{s}}(\bm X)=[\bm h_{\mathrm{s},1}(\bm X),\ldots,\bm h_{\mathrm{s},J}(\bm X)]$. 
Let $\bm a_{\mathrm{c},k}\in\cC^{1\times D}$ and $\bm a_{\mathrm{s},j}\in\cC^{1\times D}$ denote row $k$ of $\bm A_{\mathrm{c}}$ and row $j$ of $\bm A_{\mathrm{s}}$, respectively.
With $\Gamma_k=2^{R_k^{\min}}-1$, the communication rate constraint can be equivalently written as
\begin{equation}
\Omega_k(\bm a_{\mathrm{c},k})\triangleq
\Gamma_k\!\big(\norm{\bm a_{\mathrm{c},k}}_2^2+\sigma_{\mathrm{c}}^2\big)
-(1+\Gamma_k)\absval{[\bm a_{\mathrm{c},k}]_k}^2\leq0.
\label{eq:Omega}
\end{equation}
Based on this reformulation, problem~\eqref{prob:P1} is equivalently written as
\begin{align}
\min_{\bm A_{\mathrm{c}},\bm A_{\mathrm{s}},\bm B,\bm X,\bm q_{\rm U}}\quad
&\chi\sum_{j\in\cJ}\frac{\omega_j\zeta_j(\bm q_{\rm U})}{\norm{\bm a_{\mathrm{s},j}}_2^2}\label{prob:P2} \\
\mathrm{s.t.}\quad&\eqref{prob:P1_power},\ \eqref{prob:P1_X},\ \eqref{prob:P1_Q},\ \eqref{eq:auxiliary_equalities},\ \eqref{eq:Omega}.\nonumber
\end{align}
With Lagrange multipliers $\bm Y_{\mathrm{c}}\in\cC^{K\times D}$ and $\bm Y_{\mathrm{s}}\in\cC^{J\times D}$ and penalty parameter $\rho>0$, the augmented Lagrangian (AL) subproblem of~\eqref{prob:P2} is
\begin{align}
\min_{\bm A_{\mathrm{c}},\bm A_{\mathrm{s}},\bm B,\bm X,\bm q_{\rm U}}\quad & \mathcal L_{\mathrm{A}} \label{prob:P3}\\
\mathrm{s.t.}\quad&\eqref{prob:P1_power},\ \eqref{prob:P1_X},\ \eqref{prob:P1_Q},\ \eqref{eq:Omega},\nonumber
\end{align}
where
$
\mathcal L_{\mathrm{A}}
={}\chi\sum_{j\in\cJ}\frac{\omega_j\zeta_j(\bm q_{\rm U})}{\|\bm a_{\mathrm{s},j}\|_2^2}
+\frac{1}{2\rho}\|\bm A_{\mathrm{c}}-\bm H_{\mathrm{c}}^{\mathrm{H}}(\bm X)\bm B+\rho\bm Y_{\mathrm{c}}\|_{\mathrm{F}}^2
+\frac{1}{2\rho}\|\bm A_{\mathrm{s}}-\bm H_{\mathrm{s}}^{\mathrm{H}}(\bm X)\bm B+\rho\bm Y_{\mathrm{s}}\|_{\mathrm{F}}^2.
$\vspace{-0.1cm}

\subsection{Solving Problem \eqref{prob:P3}}
The primal variable blocks are updated successively, and iteration indices are omitted below for brevity.

\subsubsection{Optimization of $\bm A_{\mathrm{c}}$}
For fixed $\bm B$ and $\bm X$, row $k$ of the optimal $\bm A_{\mathrm{c}}$ solves
\begin{equation}
\min_{\bm a_{\mathrm{c},k}}\ \norm{\bm a_{\mathrm{c},k}-\bm t_{\mathrm{c},k}}_2^2\quad\mathrm{s.t.}\quad\Omega_k(\bm a_{\mathrm{c},k})\leq0,
\label{prob:Ac}
\end{equation}
where $\bm t_{\mathrm{c},k}\in\cC^{1\times D}$ is row $k$ of $\bm H_{\mathrm{c}}^{\mathrm{H}}(\bm X)\bm B-\rho\bm Y_{\mathrm{c}}$.
The $K$ subproblems are independent and can therefore be solved in parallel using the following lemma.
\begin{lemma}\label{lem:Ac_projection}
\emph{If $[\bm t_{\mathrm{c},k}]_k\neq0$, the unique global solution of \eqref{prob:Ac} is
\begin{equation}
[\bm a_{\mathrm{c},k}]_k^*=\frac{[\bm t_{\mathrm{c},k}]_k}{1-\mu_k},\qquad
[\bm a_{\mathrm{c},k}]_i^*=\frac{[\bm t_{\mathrm{c},k}]_i}{1+\mu_k\Gamma_k},\ i\neq k,
\label{eq:Ac_update}
\end{equation}
where $\mu_k=0$ if $\Omega_k(\bm t_{\mathrm{c},k})\leq0$; otherwise, $\mu_k\in(0,1)$ is the unique root of $|[\bm t_{\mathrm{c},k}]_k|^2/(1-\mu_k)^2=\Gamma_k[\sum_{i\neq k}|[\bm t_{\mathrm{c},k}]_i|^2/(1+\mu_k\Gamma_k)^2+\sigma_{\mathrm{c}}^2]$, found by bisection. If $[\bm t_{\mathrm{c},k}]_k=0$, then $[\bm a_{\mathrm{c},k}]_i=[\bm t_{\mathrm{c},k}]_i/(1+\Gamma_k)$ for $i\neq k$ and
\begin{equation}
[\bm a_{\mathrm{c},k}]_k^*=e^{\jmathc\vartheta_k}\!\left[\Gamma_k\!\left(\frac{\sum_{i\neq k}|[\bm t_{\mathrm{c},k}]_i|^2}{(1+\Gamma_k)^2}+\sigma_{\mathrm{c}}^2\right)\right]^{1/2}
\label{eq:Ac_degenerate}
\end{equation}
is optimal for any $0\le\vartheta_k<2\pi$ and we set $\vartheta_k=0$.
}
\end{lemma}
\begin{proof}
See Appendix~\ref{app:proof_Ac}.
\end{proof}

\subsubsection{Optimization of $\bm A_{\mathrm{s}}$}
For fixed $\bm B$, $\bm X$, and $\bm q_{\rm U}$, the optimization of $\mathbf A_{\mathrm{s}}$ decomposes into $J$ independent subproblems, which can be solved in parallel:
\begin{equation}
\min_{\bm a_{\mathrm{s},j}}\ \frac{\chi\omega_j\zeta_j(\bm q_{\rm U})}{\norm{\bm a_{\mathrm{s},j}}_2^2}+\frac{1}{2\rho}\norm{\bm a_{\mathrm{s},j}-\bm t_{\mathrm{s},j}}_2^2,
\quad j\in\cJ,
\label{prob:As}
\end{equation}
where $\bm t_{\mathrm{s},j}\in\cC^{1\times D}$ is row $j$ of $\bm H_{\mathrm{s}}^{\mathrm{H}}(\bm X)\bm B-\rho\bm Y_{\mathrm{s}}$. If $\bm t_{\mathrm{s},j}\neq\bm0$, the unique global solution is $\bm a_{\mathrm{s},j}=u_j\bm t_{\mathrm{s},j}/\norm{\bm t_{\mathrm{s},j}}_2$, where $u_j>0$ is the unique root of
\begin{equation}
u_j^4-\norm{\bm t_{\mathrm{s},j}}_2u_j^3-2\rho\chi\omega_j\zeta_j(\bm q_{\rm U})=0.
\label{eq:As_root}
\end{equation}
If $\bm t_{\mathrm{s},j}=\bm0$, we set $\bm a_{\mathrm{s},j}=[2\rho\chi\omega_j\zeta_j(\bm q_{\rm U})]^{1/4}\bm e_1^{\mathrm{T}}$, where $\bm e_1=[1,0,\ldots,0]^{\mathrm{T}}\in\cC^{D\times1}$.

\subsubsection{Optimization of $\bm B$}
For fixed $\bm A_{\mathrm{c}}$, $\bm A_{\mathrm{s}}$, and $\bm X$, the beamforming subproblem is
\begin{equation}
\begin{aligned}
\min_{\bm B}\quad&\norm{\bm H_{\mathrm{c}}^{\mathrm{H}}(\bm X)\bm B-\bm A_{\mathrm{c}}-\rho\bm Y_{\mathrm{c}}}_{\mathrm{F}}^2\\
&\quad+\norm{\bm H_{\mathrm{s}}^{\mathrm{H}}(\bm X)\bm B-\bm A_{\mathrm{s}}-\rho\bm Y_{\mathrm{s}}}_{\mathrm{F}}^2\\
\mathrm{s.t.}\quad&\norm{\bm B}_{\mathrm{F}}^2\leq P_{\max}.
\end{aligned}
\label{prob:B}
\end{equation}
Let $\bm C_B=\bm H_{\mathrm{c}}(\bm X)\bm H_{\mathrm{c}}^{\mathrm{H}}(\bm X)+\bm H_{\mathrm{s}}(\bm X)\bm H_{\mathrm{s}}^{\mathrm{H}}(\bm X)\in\cC^{N\times N}$ and $\bm D_B=\bm H_{\mathrm{c}}(\bm X)(\bm A_{\mathrm{c}}+\rho\bm Y_{\mathrm{c}})+\bm H_{\mathrm{s}}(\bm X)(\bm A_{\mathrm{s}}+\rho\bm Y_{\mathrm{s}})\in\cC^{N\times D}$. A global solution is
\begin{equation}
\bm B^*=(\bm C_B+\lambda_B^\star\bm I_N)^\dagger\bm D_B,
\label{eq:B_update}
\end{equation}
where $\lambda_B^\star=0$ if
$\|\bm C_B^\dagger\bm D_B\|_{\mathrm{F}}^2\le P_{\max}$.
Otherwise, the unique $\lambda_B^\star>0$ satisfying
$\|\bm B^\star\|_{\mathrm{F}}^2=P_{\max}$ is found by bisection.
Here, $(\cdot)^\dagger$ denotes the Moore-Penrose pseudoinverse. 

\subsubsection{Optimization of $\bm X$}
For fixed $\bm A_{\mathrm{c}}$, $\bm A_{\mathrm{s}}$, and $\bm B$, the PA position optimization subproblem is
\begin{equation}
\min_{\bm X\in\cX}\ g_X(\bm X)\triangleq
\norm{\bm R_{\mathrm{c}}}_{\mathrm{F}}^2+\norm{\bm R_{\mathrm{s}}}_{\mathrm{F}}^2,
\label{prob:X}
\end{equation}
where $\bm R_{\mathrm{c}}=\bm H_{\mathrm{c}}^{\mathrm{H}}(\bm X)\bm B-\bm A_{\mathrm{c}}-\rho\bm Y_{\mathrm{c}}\in\cC^{K\times D}$ and $\bm R_{\mathrm{s}}=\bm H_{\mathrm{s}}^{\mathrm{H}}(\bm X)\bm B-\bm A_{\mathrm{s}}-\rho\bm Y_{\mathrm{s}}\in\cC^{J\times D}$. Its gradient is
$
[\nabla g_X]_{m,n}=2\Real\{\Tr[\bm R_{\mathrm{c}}^{\mathrm{H}}\frac{\partial\bm H_{\mathrm{c}}^{\mathrm{H}}}{\partial x_{m,n}}\bm B]+\Tr[\bm R_{\mathrm{s}}^{\mathrm{H}}\frac{\partial\bm H_{\mathrm{s}}^{\mathrm{H}}}{\partial x_{m,n}}\bm B]\}.
$
The update at the current iterate $\bm X_0$ is given as
\begin{equation}
\bm X^{+}=\arg\min_{\bm X\in\cX}
\left\|\bm X-\bm X_0+\frac{\nabla g_X(\bm X_0)}{L_X}\right\|_{\mathrm{F}}^2.
\label{eq:X_projected}
\end{equation}
where $L_X$ is chosen by backtracking.
Let $z_{m,n}=x_{m,n}-x_{\mathrm{f}}-(m-1)\delta$ and $\widehat z_{m,n}=[\bm X_0]_{m,n}-x_{\mathrm{f}}-[\nabla g_X(\bm X_0)]_{m,n}/L_X-(m-1)\delta$. 
The projection in~\eqref{eq:X_projected} can be computed by solving the following problem independently for each waveguide $n$
\begin{equation}
\min_{0\leq z_{1,n}\leq\cdots\leq z_{M,n}\leq L-(M-1)\delta}\sum\nolimits_{m=1}^{M}(z_{m,n}-\widehat z_{m,n})^2,
\label{eq:isotonic}
\end{equation}
Problem~\eqref{eq:isotonic} is a bounded isotonic regression problem, which can be solved using the pool-adjacent-violators algorithm.
The PA positions are then recovered as $x_{m,n}^{+}=x_{\mathrm{f}}+z_{m,n}^{\star}+(m-1)\delta$.
Backtracking increases $L_X\leftarrow\tau_XL_X$, $\tau_X>1$, until $g_X(\bm X^+)\leq g_X(\bm X_0)+\langle\nabla g_X(\bm X_0),\bm X^+-\bm X_0\rangle+L_X\norm{\bm X^+-\bm X_0}_{\mathrm{F}}^2/2$.

\subsubsection{Optimization of $\bm q_{\rm U}$}
{
For fixed $\bm A_{\mathrm{s}}$, the UAV position subproblem is
\begin{equation}
\min_{\bm q_{\rm U}\in\cQ}\ f_Q(\bm q_{\rm U})\triangleq
\chi\sum_{j\in\cJ}\frac{\omega_j\zeta_j(\bm q_{\rm U})}{\norm{\bm a_{\mathrm{s},j}}_2^2}.
\label{prob:Q}
\end{equation}
For potential position $j$, let $\Delta_{x,j}=x_{\mathrm{s},j}-x_{\rm U}$, $\Delta_{y,j}=y_{\mathrm{s},j}-y_{\rm U}$, $q_j=\Delta_{y,j}^2+h_{\rm U}^2$, and $d_j^2=\Delta_{x,j}^2+q_j$.
Then $\nabla_{x,y}f_Q=\chi\sum_{j\in\cJ}\omega_j\nabla_{x,y}\zeta_j/\norm{\bm a_{\mathrm{s},j}}_2^2$, where $\zeta_j=d_j^4/q_j$, and $\nabla_{x,y}\zeta_j=[-4d_j^2\Delta_{x,j}/q_j,\allowbreak\,2d_j^2\Delta_{y,j}(d_j^2-2q_j)/q_j^2]^{\mathrm{T}}$.
Since $h_{\rm U}$ is fixed, define $\nabla f_Q=[(\nabla_{x,y}f_Q)^{\mathrm{T}},0]^{\mathrm{T}}$.
The candidate update is
\begin{equation}
\bm q_{\rm U}^{+}
=\Pi_{\cQ}\!\left(
\bm q_{\rm U}-\frac{\nabla f_Q(\bm q_{\rm U})}{L_Q}
\right),
\label{eq:Q_update}
\end{equation}
where $\Pi_{\cQ}$ denotes Euclidean projection onto $\cQ$.
Backtracking increases $L_Q\leftarrow\tau_QL_Q$, with $\tau_Q>1$, and recomputes $\bm q_{\rm U}^{+}$ until
$
f_Q(\bm q_{\rm U}^{+})
\leq f_Q(\bm q_{\rm U})
+\langle\nabla f_Q(\bm q_{\rm U}),\bm q_{\rm U}^{+}-\bm q_{\rm U}\rangle
+\frac{L_Q}{2}\norm{\bm q_{\rm U}^{+}-\bm q_{\rm U}}_2^2.
$
The update is then accepted by setting $\bm q_{\rm U}\leftarrow\bm q_{\rm U}^{+}$.
\par}\vspace{-0.1cm}

\subsection{Overall Solution and Complexity}
The overall algorithm is summarized in Algorithm~\ref{alg:PDD}, where $t$ and $r$ denote the outer and inner iteration indices. 
Define $\bm E_{\mathrm{c}}=\bm A_{\mathrm{c}}-\bm H_{\mathrm{c}}^{\mathrm{H}}(\bm X)\bm B$ and $\bm E_{\mathrm{s}}=\bm A_{\mathrm{s}}-\bm H_{\mathrm{s}}^{\mathrm{H}}(\bm X)\bm B$. At outer iteration $t$, the constraint violation is $\Lambda^{(t)}=\max\{\|\bm E_{\mathrm{c}}^{(t)}\|_{\mathrm{F}},\|\bm E_{\mathrm{s}}^{(t)}\|_{\mathrm{F}}\}$, and the relative AL change in the inner loop is $\Delta_{\mathrm{A}}^{(t,r)}=|\mathcal L_{\mathrm{A}}^{(t,r)}-\mathcal L_{\mathrm{A}}^{(t,r-1)}|/|\mathcal L_{\mathrm{A}}^{(t,r-1)}|$.
At each outer iteration, the primal variable blocks are iteratively updated until $\Delta_{\mathrm{A}}^{(t,r)}\leq\epsilon^{(t)}$ or $r=I_{\rm in}^{\max}$, where $\epsilon^{(t+1)}=c_\epsilon\epsilon^{(t)}$ with $0<c_\epsilon<1$. 
If $\Lambda^{(t+1)}\leq\eta^{(t)}$, the multipliers are updated as $\bm Y_i^{(t+1)}=\bm Y_i^{(t)}+\bm E_i^{(t+1)}/\rho^{(t)}$, $i\in\{\mathrm{c},\mathrm{s}\}$, and $\eta^{(t+1)}=c_\eta\eta^{(t)}$, where $0<c_\eta<1$.
Otherwise, $\rho^{(t+1)}=c_\rho\rho^{(t)}$, where $0<c_\rho<1$. 
The outer loop terminates when $\Lambda^{(t)}\leq\epsilon_{\rm out}$ or $t=I_{\rm out}^{\max}$. 
Under mild conditions, every accumulation point of the sequence generated by Algorithm~\ref{alg:PDD} is a KKT point of problem~\eqref{prob:P2}~\cite{shi2020pdd}.

Let $I_{\rm out}$ denote the actual number of outer iterations and $\overline I_{\rm in}$ the average number of inner iterations per outer iteration.
Let $\overline I_\mu$ and $\overline I_\lambda$ denote the average numbers of bisection steps for updating $\bm A_{\mathrm{c}}$ and $\bm B$, respectively, and let $\overline I_X$ and $\overline I_Q$ denote the average numbers of backtracking steps for updating $\bm X$ and $\bm q_{\rm U}$, respectively.
The overall computational complexity is $\mathcal O\!\bigl(I_{\rm out}\overline I_{\rm in}[KD \overline I_\mu+JD+N^3 \overline I_\lambda+MN \overline I_X(K+J)D  +J \overline I_Q]\bigr)$.\vspace{-0.1cm}

\begin{algorithm}[t]
\footnotesize
\caption{Proposed PDD Algorithm}
\label{alg:PDD}
\begin{algorithmic}[1]
\REQUIRE $\rho^{(0)}>0$, $\epsilon^{(0)}>0$, $\eta^{(0)}>0$, $c_\rho,c_\epsilon,c_\eta\in(0,1)$, $\epsilon_{\rm out}>0$, $I_{\rm in}^{\max}$, and $I_{\rm out}^{\max}$.
\STATE Initialize $\bm X$, $\bm B$, $\bm q_{\rm U}$, $\bm A_{\mathrm{c}}$, $\bm A_{\mathrm{s}}$, $\bm Y_{\mathrm{c}}$, and $\bm Y_{\mathrm{s}}$.
\STATE Set $t\leftarrow0$.
\REPEAT
    \STATE Set $r\leftarrow0$ and evaluate $\mathcal L_{\mathrm{A}}^{(t,0)}$.
    \REPEAT
        \STATE Sequentially update $\bm A_{\mathrm{c}}$, $\bm A_{\mathrm{s}}$, $\bm B$, $\bm X$, and $\bm q_{\rm U}$.
        \STATE $r\leftarrow r+1$.
    \UNTIL{$\Delta_{\mathrm{A}}^{(t,r)}\leq\epsilon^{(t)}$} or $r=I_{\rm in}^{\max}$
    \STATE Compute $\Lambda^{(t+1)}$.
    \IF{$\Lambda^{(t+1)}\leq\eta^{(t)}$}
        \STATE Set $\bm Y_i^{(t+1)}\leftarrow\bm Y_i^{(t)}+\bm E_i^{(t+1)}/\rho^{(t)}$, $i\in\{\mathrm{c},\mathrm{s}\}$; set $\rho^{(t+1)}\leftarrow\rho^{(t)}$ and $\eta^{(t+1)}\leftarrow c_\eta\eta^{(t)}$.
    \ELSE
        \STATE Retain $\bm Y_{\mathrm{c}}$ and $\bm Y_{\mathrm{s}}$; set $\rho^{(t+1)}\leftarrow c_\rho\rho^{(t)}$ and $\eta^{(t+1)}\leftarrow\eta^{(t)}$.
    \ENDIF
    \STATE $\epsilon^{(t+1)}\leftarrow c_\epsilon\epsilon^{(t)}$ and $t\leftarrow t+1$.
\UNTIL{$\Lambda^{(t)}\leq\epsilon_{\rm out}$ or $t=I_{\rm out}^{\max}$}
\RETURN $\bm B$, $\bm X$, and $\bm q_{\rm U}$.
\end{algorithmic}
\end{algorithm}

\section{Numerical Results}\label{sec:numerical_results}
\subsection{Simulation Setup}
We consider a ground service area $[5,35]~\mathrm{m}\times[0,30]~\mathrm{m}$, with $x_{\mathrm{f}}=5~\mathrm{m}$. 
The $K=5$ user locations and the target location are uniformly sampled over the service area.
The $J=4$ candidate target locations lie within a $2~\mathrm{m}\times2~\mathrm{m}$ area, whose center is uniformly randomly selected from the feasible locations within the ground service area, with $\omega_j=1/J$ for all $j\in\cJ$.
The BS is deployed at $(0,15,5)~{\rm m}$, and is equipped with a PASS comprising $N=6$ waveguides, each carrying $M=4$ PAs over a length of $L=30~{\rm m}$.
The waveguide coordinates are $y_n=30n/(N+1)~\mathrm{m}$, $n=1,\ldots,N$.
The UAV is equipped with an $8$-element ULA and has its horizontal position optimized within the service area at a fixed altitude of $h_{\rm U}=15~{\rm m}$. 
We set $f_{\mathrm{c}}=28~{\rm GHz}$, $\sigma_{\rm RCS}=0.1~\mathrm{m}^2$, $\alpha=\sqrt{4\pi\sigma_{\rm RCS}}/\lambda_{\mathrm{c}}$, $\delta=\lambda_{\mathrm{c}}/2$, $n_{\mathrm{e}}=1.4$, $d_{\rm U}=\lambda_{\mathrm{c}}/2$, $P_{\max}=30~{\rm dBm}$, $\Gamma_k=6~{\rm dB}$, $\sigma_{\mathrm{c}}^2=\sigma_{\mathrm{s}}^2=-90~{\rm dBm}$, and $T_{\mathrm{s}}=256$.
The PDD parameters are $\rho^{(0)}=1$, $c_\rho=0.5$, $\epsilon^{(0)}=10^{-2}$, $c_\epsilon=0.95$, $\eta^{(0)}=10^{-6}$, $c_\eta=0.9$, and $\epsilon_{\rm out}=10^{-8}$, $I_{\rm in}^{\max}=100$, $I_{\rm out}^{\max}=100$, $L_X^{(0)}=L_Q^{(0)}=1$, and $\tau_X=\tau_Q=2$. 
All results are averaged over $50$ realizations.
For comparison, we consider the following independently optimized benchmark schemes: 1) \emph{conventional MIMO}, which employs an $N$-antenna fully digital array; 2) \emph{massive MIMO}, which employs partially connected hybrid beamforming with $MN$ antennas and $N$ RF chains; 3) \emph{uniform PASS}, where the PAs are uniformly fixed along each waveguide; 4) \emph{discrete PASS}, where each PA position is selected from $N_{\mathrm{d}}+1$ uniformly spaced positions in $[x_{\mathrm{f}},x_{\mathrm{f}}+L]$; 5) \emph{PASS with BS sensing}, where sensing is performed by a fixed receiver at the BS without the UAV; and 6) \emph{PASS with one-dimensional PA search}, where the PA coordinates are sequentially updated using one-dimensional search.\vspace{-0.1cm}

\subsection{Numerical Results}

\begin{figure}[!ht]
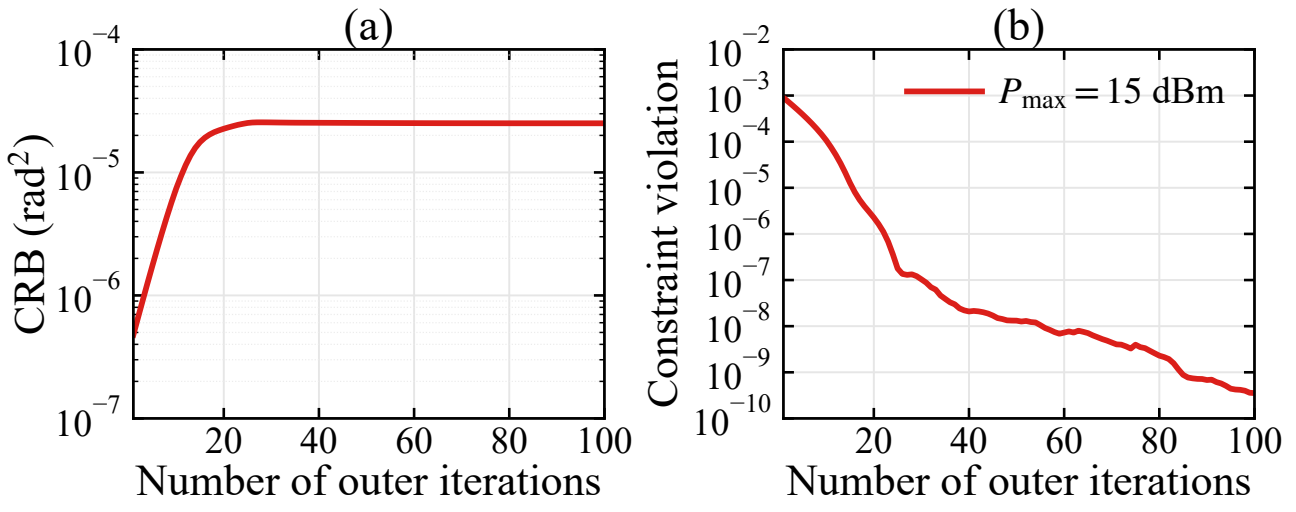

    \centering
    \safeincludegraphics[width=0.47\textwidth]{Convergence_single1.pdf}
    \caption{weighted CRB and constraint violation versus the PDD outer iteration.}\vspace{-0.2cm}
    \label{fig:convergence}
\end{figure}

Fig.~\ref{fig:convergence} plots the weighted CRB and constraint violation at $P_{\max}=15~\mathrm{dBm}$.
The weighted CRB increases during the first $20$ outer iterations and changes little after about $30$ iterations, while the constraint violation continues to decrease.

\begin{figure}[!ht]
    \centering
    \safeincludegraphics[width=0.48\textwidth]{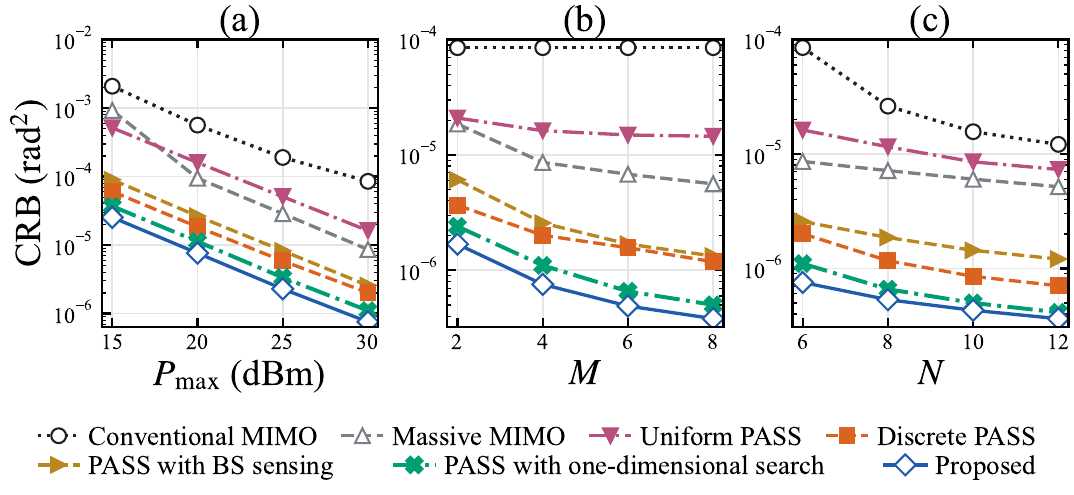}
    \caption{weighted CRB versus (a) transmit power budget $P_{\max}$, (b) number of PAs per waveguide $M$, and (c) number of waveguides $N$.}\vspace{-0.2cm}
    \label{fig:performance}
\end{figure}

Fig.~\ref{fig:performance}(a) shows that the weighted CRB decreases with $P_{\max}$ for all schemes, since a larger transmit-power budget allows more power to be used for sensing under the same rate requirements. 
The proposed scheme achieves the lowest weighted CRB through joint beamforming, PA-position, and UAV-position optimization. 
The performance gains over \emph{Discrete PASS} and \emph{PASS with one-dimensional search} reflect the limitations of discrete PA placement and one-dimensional position search, respectively. 
The performance gains over \emph{Uniform PASS} and \emph{PASS with BS sensing} demonstrate the benefits of optimizing the PA positions and using a UAV as the sensing receiver, respectively.
Compared with the MIMO baselines, the optimized PASS schemes further benefit from the adjustable PA locations at the transmitter.

Fig.~\ref{fig:performance}(b) shows that increasing the number of PAs per waveguide $M$ generally reduces the weighted CRB for the optimized PASS schemes, as it provides greater flexibility in pinching beamforming.
Although \emph{Uniform PASS} and \emph{Discrete PASS} also benefit from increasing $M$, their performance improvement is more limited due to their limited PA deployments.

Fig.~\ref{fig:performance}(c) shows that the weighted CRB decreases as $N$ increases for all schemes. 
With more waveguides, more RF chains are available at the BS, providing more degrees of freedom for beamforming.

\section{Conclusion}
This paper proposed a PASS-assisted air-ground ISAC system with a UAV sensing receiver. 
A PDD algorithm was developed to minimize the weighted CRB by jointly optimizing the PASS transmit design and UAV position.
The auxiliary variable and beamforming subproblems were shown to admit globally optimal solutions, while the PA and UAV positions were updated using SCA.
Moreover, numerical results showed that jointly optimizing the PASS transmit design and UAV position provides consistent sensing gains over the considered benchmarks.

\appendices\vspace{-0.2cm}
\section{Derivation of \eqref{eq:conditional_AoA_CRB}}\label{app:AoA_CRB}

For a potential target location $\bm l_{\mathrm{s},j}$, stacking the $T_{\mathrm{s}}$ received 
snapshots in~\eqref{eq:uav_snapshot} gives 
$\bm Y_j=\eta_j\bm a_{\rm U}(\theta_j)\bm h_{\mathrm{s},j}^{\mathrm{H}}(\bm X)\bm S+\bm N_j$, 
where $\bm Y_j=[\bm y_{{\rm U},j}[1],\ldots,\bm y_{{\rm U},j}[T_{\mathrm{s}}]]$ and 
$\bm N_j=[\bm n_{{\rm U},j}[1],\ldots,\bm n_{{\rm U},j}[T_{\mathrm{s}}]]$. Hence, 
$\widetilde{\bm y}_j\triangleq\vect(\bm Y_j)
\sim\CN(\bm\mu_j,\sigma_{\mathrm{s}}^2\bm I_{UT_{\mathrm{s}}})$, where 
$\bm\mu_j=\eta_j\vect\!\left(\bm a_{\rm U}(\theta_j)
\bm h_{\mathrm{s},j}^{\mathrm{H}}(\bm X)\bm S\right)$.

Treating $\eta_j$ as an unknown complex nuisance parameter, define 
$\bm\xi_j=[\theta_j,\Real\{\eta_j\},\Imag\{\eta_j\}]^{\mathrm{T}}$. The corresponding 
Fisher information matrix is 
$\bm J_j=(2/\sigma_{\mathrm{s}}^2)\Real\{(\partial\bm\mu_j/\partial\bm\xi_j)^{\mathrm{H}}
(\partial\bm\mu_j/\partial\bm\xi_j)\}$. The three columns of 
$\partial\bm\mu_j/\partial\bm\xi_j$ are 
$\eta_j\vect\!\left(\dot{\bm a}_{\rm U}(\theta_j)
\bm h_{\mathrm{s},j}^{\mathrm{H}}(\bm X)\bm S\right)$, 
$\vect\!\left(\bm a_{\rm U}(\theta_j)\bm h_{\mathrm{s},j}^{\mathrm{H}}(\bm X)\bm S\right)$, 
and $\jmathc\vect\!\left(\bm a_{\rm U}(\theta_j)
\bm h_{\mathrm{s},j}^{\mathrm{H}}(\bm X)\bm S\right)$, respectively, where 
$\dot{\bm a}_{\rm U}(\theta_j)
=\partial\bm a_{\rm U}(\theta_j)/\partial\theta_j$. The conditional AoA CRB 
is $C_{\theta,j}=[\bm J_j^{-1}]_{1,1}$.

Applying the Schur complement with respect to $\Real\{\eta_j\}$ and 
$\Imag\{\eta_j\}$ yields
\begin{equation}
\begin{aligned}
C_{\theta,j}
=\Bigg[
&\frac{2|\eta_j|^2}{\sigma_{\mathrm{s}}^2}
\bm h_{\mathrm{s},j}^{\mathrm{H}}(\bm X)\bm S\bm S^{\mathrm{H}}\bm h_{\mathrm{s},j}(\bm X)\\
&\times\left(
\norm{\dot{\bm a}_{\rm U}(\theta_j)}_2^2
-\frac{
\left|\bm a_{\rm U}^{\mathrm{H}}(\theta_j)
\dot{\bm a}_{\rm U}(\theta_j)\right|^2
}{
\norm{\bm a_{\rm U}(\theta_j)}_2^2
}
\right)
\Bigg]^{-1}.
\end{aligned}
\label{eq:crb_schur}
\end{equation}

For the UAV ULA, 
$\dot{\bm a}_{\rm U}(\theta_j)
=-\jmathc\kappa_{\rm U}\cos\theta_j\bm D_U\bm a_{\rm U}(\theta_j)$, 
where $\bm D_U=\diag(0,1,\ldots,U-1)$. Consequently,
$\norm{\dot{\bm a}_{\rm U}(\theta_j)}_2^2
-\left|\bm a_{\rm U}^{\mathrm{H}}(\theta_j)\dot{\bm a}_{\rm U}(\theta_j)\right|^2/
\norm{\bm a_{\rm U}(\theta_j)}_2^2
=\kappa_{\rm U}^2\cos^2\theta_j\,U(U^2-1)/12$. Moreover, using 
$T_{\mathrm{s}}^{-1}\bm S\bm S^{\mathrm{H}}\simeq\bm B\bm B^{\mathrm{H}}$, we have 
$\bm h_{\mathrm{s},j}^{\mathrm{H}}(\bm X)\bm S\bm S^{\mathrm{H}}\bm h_{\mathrm{s},j}(\bm X)
\simeq T_{\mathrm{s}}\bm h_{\mathrm{s},j}^{\mathrm{H}}(\bm X)\bm B\bm B^{\mathrm{H}}\bm h_{\mathrm{s},j}(\bm X)$. 
Substituting these results into~\eqref{eq:crb_schur} yields 
\eqref{eq:conditional_AoA_CRB}.

\section{Proof of Lemma~\ref{lem:Ac_projection}}\label{app:proof_Ac}
Let $a_i=[\bm a_{\mathrm{c},k}]_i$ and $t_i=[\bm t_{\mathrm{c},k}]_i$. The real form of \eqref{prob:Ac} has one quadratic constraint and is strictly feasible at $a_i=0$ for $i\neq k$ and $|a_k|>\sqrt{\Gamma_k}\sigma_{\mathrm{c}}$; the S-lemma thus gives strong duality, so the KKT solution is global. Stationarity gives $(1-\mu_k)a_k=t_k$ and $(1+\mu_k\Gamma_k)a_i=t_i$ for $i\neq k$, with $0\leq\mu_k\leq1$. If $\Omega_k(\bm t_{\mathrm{c},k})\leq0$, then $\mu_k=0$. Otherwise, for $t_k\neq0$, $0<\mu_k<1$ and the active constraint gives the root equation in Lemma~\ref{lem:Ac_projection}. Its left side minus the right side is strictly increasing from a negative value to $+\infty$, so the root is unique. For $t_k=0$, any $\mu_k<1$ violates feasibility; hence, $\mu_k=1$, and stationarity with the active constraint yields \eqref{eq:Ac_degenerate}, with arbitrary phase.

\bibliographystyle{IEEEtran}
\bibliography{reference}

\end{document}